\documentclass[twocolumn,prl,superscriptaddress]{revtex4-1}
\usepackage{graphicx}
\usepackage{amssymb,amsmath}
\usepackage{bm}
\usepackage{dcolumn}
\usepackage{subfigure}
\usepackage{float}
\usepackage[OT1]{fontenc} 
\usepackage{url}
\usepackage{slashed}
\usepackage{color}
\usepackage{verbatim}
\usepackage{enumitem}
\usepackage{txfonts}
\usepackage{soul,physics}
\usepackage[driverfallback=dvipdfm]{hyperref}
\hypersetup{pdfpagemode=FullScreen,colorlinks=true,breaklinks,urlcolor=blue,linkcolor=blue,citecolor=blue}
\usepackage{natbib}

\usepackage{subfigure}
\usepackage{comment}
\usepackage{hyperref}
\usepackage{amssymb}
\usepackage{bm}
\usepackage{graphicx}
\usepackage{amsmath,amssymb}
\usepackage{tikz,fp}
\usepackage{tikz-cd}
\usetikzlibrary{arrows}
\usetikzlibrary{intersections}
\usetikzlibrary{shapes.geometric}
\usetikzlibrary{decorations.pathmorphing, patterns,shapes,fixedpointarithmetic}
\usetikzlibrary{decorations.markings}

\pgfdeclarepatternformonly{south west lines}{\pgfqpoint{-0pt}{-0pt}}{\pgfqpoint{3pt}{3pt}}{\pgfqpoint{3pt}{3pt}}{
	\pgfsetlinewidth{0.4pt}
	\pgfpathmoveto{\pgfqpoint{0pt}{0pt}}
	\pgfpathlineto{\pgfqpoint{3pt}{3pt}}
	\pgfpathmoveto{\pgfqpoint{2.8pt}{-.2pt}}
	\pgfpathlineto{\pgfqpoint{3.2pt}{.2pt}}
	\pgfpathmoveto{\pgfqpoint{-.2pt}{2.8pt}}
	\pgfpathlineto{\pgfqpoint{.2pt}{3.2pt}}
	\pgfusepath{stroke}}

\tikzset{
	mid arrow/.style={postaction={decorate,decoration={
				markings,
				mark=at position .575 with {\arrow{stealth}}
	}}},
	near arrow/.style={postaction={decorate,decoration={
				markings,
				mark=at position .275 with {\arrow{stealth}}
	}}},
	far arrow/.style={postaction={decorate,decoration={
				markings,
				mark=at position .800 with {\arrow{stealth}}
	}}},
	snake arrow/.style={fixed point arithmetic, decorate, decoration={snake,amplitude=2pt, segment length=11pt},postaction={decoration={markings,mark=at position 0.625 with {\arrow{stealth}}},decorate}},
}
\tikzset{
  baseline = -0.5ex,
  wavy/.style = {
    thick,
    decorate,
    decoration={snake,amplitude=2pt,segment length=5pt}},
  sdot/.style = {
    circle,
    draw=none,
    fill=black,
    minimum size=2.5pt,
    inner sep=0pt},
  bdot/.style = {
    circle,
    draw=none,
    fill=black,
    minimum size=4pt,
    inner sep=0pt},
  svertex/.style = {
    circle,
    draw=black,
    thick,
    fill=lightgray,
    minimum size=8pt,
    inner sep=1pt},
  bvertex/.style = {
    circle,
    draw=black,
    thick,
    fill=lightgray,
    minimum size=24pt},
  bvertexsmall/.style = {
    circle,
    draw=black,
    thick,
    fill=lightgray,
    minimum size=7pt},
  bvertexnormal/.style = {
    circle,
    draw=black,
    thick,
    fill=lightgray,
    minimum size=16pt},
  dvertex/.style = {
    circle,
    draw=black,
    thick,
    fill=gray,
    minimum size=25pt}}

\begin{document}
	
	\title{Quantized Topological Response in Trapped Quantum Gases}
	
	\author{Pengfei Zhang}
	\affiliation{Institute for Quantum Information and Matter \& Walter Burke Institute for Theoretical Physics, California Institute of Technology, Pasadena, CA 91125, USA}
	\begin{abstract}
    In this letter, we propose a quantized topological response in trapped 1D quantum gases. The experimental protocol for the response requires the application of an instant optical pulse to a half-infinite region in an asymptotically harmonic trap and measuring the density distribution. We show that the corresponding linear response is described by a universal quantized formula in the thermal dynamical limit, which is invariant under local continuous deformations of the trapping potential $V$, atom distribution $f_\Lambda$, the spatial envelope of the optical pulse $\Theta_p$, and the measurement region $\Theta_m$. We test the statement by various numerical analysis, the result of which is consistent with the analytical prediction to high accuracy. We further show that a short but finite optical pulse duration only results in a violation of the quantization near the transition time, which suggests that quantized response could be observed in realistic experiments. We also generalize our results to  non-linear quantized topological responses for atoms in higher dimensional harmonic traps. 
	\end{abstract}
	
	\maketitle

    \emph{\color{blue}Introduction.--} Understanding phases of quantum many-body systems is one of the most important subjects in condensed matter physics. Nowadays, it has been realized that quantum systems are not only classified by their local order parameters \cite{landau2013statistical}, but also by their topological properties \cite{RevModPhys.82.3045,RevModPhys.83.1057,bernevig2013topological,Witten:2015aoa,RevModPhys.89.041004,RevModPhys.90.015001,moessner2021topological,zeng2019quantum,KITAEV20062,RevModPhys.88.021004,goldman2016topological}. By definition, the topological properties of quantum systems are invariant under continuous deformations, and thus much more stable against small perturbations. In certain cases, non-trivial quantum topology implies novel quantum responses, which can be directly measured in both solid-state materials \cite{PhysRevLett.45.494,doi:10.1126/science.1105514,doi:10.1126/science.1234414} and quantum simulators \cite{jotzu2014experimental,tarnowski2019measuring,PhysRevLett.121.250403,nakajima2016topological,aidelsburger2015measuring,atala2014observation}. As an example, without any symmetry restriction, band insulators in 2D can be classified by the Chern number of occupied bands \cite{PhysRevLett.49.405,PhysRevLett.93.206602}. A non-zero Chern number guarantees the existence of chiral edges states under the open boundary condition, which contributes to quantized Hall conductance. Adding symmetry constrains further lead to the new concept of symmetry protected topological phases, examples of which include the celebrated quantum spin Hall effect \cite{PhysRevLett.95.146802,PhysRevLett.96.106802,doi:10.1126/science.1148047}. Unveiling new topological responses beyond the current knowledge is then of special interests.

    Recently, C. L. Kane proposed the topology of the Fermi surface implies a novel quantized non-linear charge transport in $D$-dimension with $D\geq 2$ \cite{PhysRevLett.128.076801}, which is a generalization of the Landauer formula in 1D \cite{doi:10.1063/1.531590,PhysRevB.23.6851,Wharam_1988,Honda_1995,van2013quantized,doi:10.1126/science.280.5370.1744,krinner2015observation,krinner2017two,PhysRevLett.123.193605}. Studies further show that this Fermi surface topology can also be detected in the entanglement entropy \cite{tam2022topological}. Later, a concrete experimental protocol for observing such quantized non-linear transport has been proposed in a pioneered work \cite{yang2022quantized}, where authors study the non-interacting Fermi gases in 2D traps using semi-classical Boltzmann equations. For harmonic traps, a closed-form expression is obtained, which shows a quantization for arbitrary evolution time, with repeated transitions of the quantization value. In particular, this quantization goes beyond the early-time regime determined by the Fermi surface topology without the trapping potential. However, it is possible that such a quantization replies on strictly harmonic traps and details of the experimental protocol, which makes the underlying physics less universal. 

    \begin{figure}[t]
\centering
\includegraphics[width=0.98\linewidth]{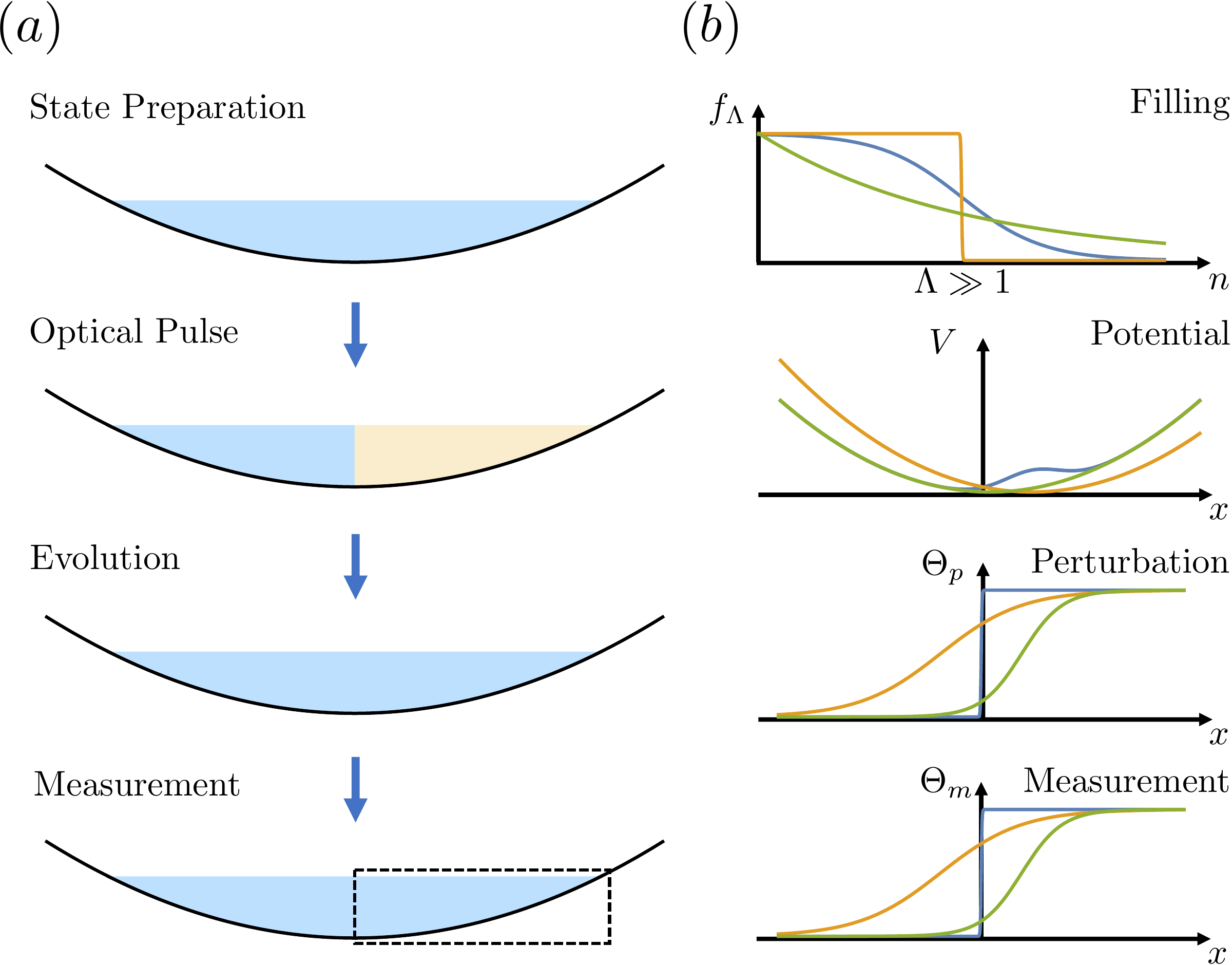}
\caption{(a). A sketch of the experimental protocol for the quantized topological response. Details of the protocol are given in the main text. (b). Some examples for local continuous deformations of $(f_\Lambda, V, \Theta_p, \Theta_m)$ with fixed boundary conditions under which the universal quantized response \eqref{eqn:statement} is valid. }\label{fig1}
\end{figure}
    In this letter, we show that it is indeed the opposite: The quantization for arbitrary time $t$ is stable against local continuous deformations of both trapping potential and experimental details as illustrated in FIG \ref{fig1}, and thus defines a new variant of universal topological response in trapped quantum many-body systems. As we will explain, the non-linear quantized response in $D$-dimensional harmonic traps is a direct consequence of its 1D counterparts. Consequently, we will focus on the 1D case in the main part of the letter. Below, we will firstly state our main conclusion in 1D, with a proposal of the experimental protocol, and then turn to the technique proofs supported by various numerical results. We also discuss the practical considerations that are necessary for realistic experiments as in \cite{yang2022quantized}. Finally, we will explain the generalization to higher dimensions, which transforms our analysis in 1D to make predictions for higher dimensions. Our theory can be tested in the near-term experiments using ultracold atomic gases.

    \vspace{5pt}

    \emph{\color{blue}The statement.--} We first state our main result in 1D: Let us consider quantum systems described by the 1D single-particle Hamiltonian
    \begin{equation}
    \hat{H}=\frac{\hat p^2}{2m}+V(\hat{x}),
    \end{equation}
    with asymptotically quadratic trapping potential $V(x)\rightarrow \frac{1}{2}m\omega^2 x^2$ for $|x|\rightarrow \infty$ \footnote{In realistic experiments, the trapping potential always deviates from being perfect harmonic at a energy scale $E_0$. Then, we should require that $E_0\gg \Lambda\gg 1$.}. For conciseness, we set $m=\omega=1$ throughout the manuscript. We label the single-particle eigenstate of $\hat H$ with energy $E_n$ as $|n\rangle$. Then the main conclusion of this letter reads
    \begin{equation}\label{eqn:statement}
    \mathcal{P}=\lim_{\Lambda \rightarrow \infty}2\pi i\sum_n f_\Lambda(n)\langle n| [\hat U^\dagger \Theta_m(\hat{x})\hat U,\Theta_p(\hat{x})]|n\rangle=\text{sgn}(\sin t).
    \end{equation}
    Here $\hat{U}=e^{-i\hat {H}t}$ and we fix the convention that $\text{sgn}(0)=0$. $f_\Lambda(n)$ is a regulator at the energy scale $\Lambda$, with $f_\Lambda(n\ll \Lambda)=1$ and $f_\Lambda(n\gg \Lambda)=0$. Physically, it describes the filling fraction of each state, and the limit $\Lambda \rightarrow \infty$ is equivalent to the thermodynamical limit, where the system contains large number of atoms. Similarly $\Theta_{m/p}(x)$ are functions that satisfy the boundary condition 
    \begin{equation}
    \Theta_{m/p}(x)=\begin{cases}
    1 & \quad x\rightarrow \infty, \\
    0 & \quad x\rightarrow -\infty.
    \end{cases}
    \end{equation}   
    Since the R.H.S. of \eqref{eqn:statement} is independent of the details of $(f_\Lambda,V,\Theta_p,\Theta_m)$, the quantized number $\mathcal{P}$ is topological, which means the invariance under local continuous deformations.

    Before getting into technical details for the proof of \eqref{eqn:statement}, let us first discuss the experimental relevance of the statement. \eqref{eqn:statement} is a summation over retarded Green's functions of single-particle states $|n\rangle $. Consequently, it is naturally related to the linear response of non-interacting many-body systems. The experimental protocol for verifying our statement contains 4 steps as sketched in FIG \ref{fig1}, which is similar to the 2D protocol proposed in \cite{yang2022quantized}:
    \begin{enumerate}
    \item We prepare many-body system, where the filling fraction of each single-particle state $|n\rangle $ is given by $f_n=f_\Lambda(n)$. An example is the thermal equilibrium state of Fermi gases with $f_\Lambda(n)=(e^{\beta (E_n-\Lambda)}+1)^{-1}$. Here $\Lambda$ plays the role of the chemical potential.

    \item We apply an optical pulse to create a potential \cite{zhai2021ultracold} $\hat V_p=\xi \Theta_p(\hat{x})\delta(t)$ for a half-infinite region. After the pulse, each single-particle state becomes $|\psi_n(0^+)\rangle=e^{-i\xi \Theta_p(\hat x)}|n\rangle$.

    \item The density of the system $\rho(x,\xi)$ is measured after the evolution of time $t$. We then analyze the experimental data by computing $\rho_\xi=\int dx~\rho(x,\xi)\Theta_m(x)$. 

    \item We repeat steps 1-3 for different $\xi$ and extract the response at small $\xi$. In the thermodynamical limit $\Lambda \gg 1$, the statement \eqref{eqn:statement} predicts a quantized response $$-2\pi\left.\partial_\xi \rho_\xi \right|_{\xi=0}=\mathcal{P}=\text{sgn}(\sin t).$$ 

    \end{enumerate}

    Now we turn to the proof of the statement. In the following sections, we prove \eqref{eqn:statement} by firstly evaluating the L.H.S. using a particular choice of $(f_\Lambda, V, \Theta_p, \Theta_m)$, and then revealing its topological nature by showing its invariance under local continuous deformations of these functions. Direct numerical verifications will also be presented. 
    \vspace{5pt}

   \emph{\color{blue}Explicit calculation.--} We first compute \eqref{eqn:statement} in a particular setup with $f_\Lambda(n)=e^{-E_n/\Lambda}$ and $V(x)=\frac{1}{2} x^2$. Introducing $\epsilon=\Lambda^{-1}$, the L.H.S. of \eqref{eqn:statement} becomes
   \begin{equation}\label{eqn:explicit1}
   \mathcal{P}=4\pi\lim_{\epsilon \rightarrow 0}\text{Im}~\text{tr}\left[e^{-\epsilon \hat{H}}\Theta_p(\hat{x})e^{i\hat Ht} \hat \Theta_m(\hat{x}) e^{-i\hat Ht}\right].
   \end{equation}
   Here the trace is over the single-particle Hilbert space. In our choice, the regulator becomes a small imaginary time evolution, which makes an explicit calculation possible. Using the single-particle Green's function $K(x,y,t)=\langle x|e^{-i\hat H t}|y\rangle$, we can write \eqref{eqn:explicit1} as 
   \begin{equation}\label{eqn:harmonicgen}
   \mathcal{P}=4\pi\lim_{\epsilon \rightarrow 0}\text{Im}\int dxdy~\Theta_m(x)\Theta_p(y)K(x,y,t-i\epsilon)K(y,x,-t).
   \end{equation}
   In harmonic traps, we have a closed-form expression \cite{altland2010condensed} $K(x,y,t)=\frac{1}{\sqrt{2\pi i \sin t}}\exp(\frac{i}{2\sin t}\Big[(x^2+y^2)\cos t-2xy\Big]).$ We further choose $\Theta_p(x)=\Theta_m(x)=\theta(x)$. Here $\theta(x)$ is the unit step function. For $t\neq n\pi$, we can perform the integral over $x$ and $y$, and expand for small $\epsilon$. Leaving details into the supplementary material \cite{SM}, we find
   \begin{equation}\label{eqn:explicit2}
   \mathcal{P}=4\pi\lim_{\epsilon \rightarrow 0}\left(\frac{|\sin t|}{4\pi \sin t}+O(\epsilon^2)\right)=\text{sgn}(\sin t).
   \end{equation}
   This also indicates the result \eqref{eqn:statement} converges with power-law corrections for smooth cutoff function $f_\Lambda(n)$. We also need to examine results at the transition time $t= n\pi$. In this case, the Green's function is proportional to $\delta(x-(-1)^ny)$, which is equivalent to the identity operator $\hat I$ or the parity operator $\hat P$. In either case, $\hat U^\dagger \Theta_m(\hat{x})\hat U$ is then diagonal in real-space, and thus commutes with $\Theta_p(\hat{x})$. 
    \vspace{5pt}

   \begin{table}[t]
   \begin{tabular}{|l|l|l|l|l|}
   \hline
    & $t=\frac{\pi}{3}$ &$t=\frac{\pi}{2}$  & $t=\frac{3\pi}{2}$ & $t=\frac{7\pi}{4}$ \\ \hline
    $\Theta_m=\theta(x),\ \Theta_p=\theta(x)$& 1.0000 & 1.0000 & -1.0000 & -1.0000 \\ \hline
    $\Theta_m=\theta(x+1),\ \Theta_p=\theta(x+1)$& 0.9999 & 0.9999 & -0.9999 & -0.9999 \\ \hline
    $\Theta_m^{-1}=(e^{-x}+1),\ \Theta_p^{-1}=(e^{-x}+1)$& 0.9996  & 0.9997&  -0.9997& -0.9993 \\ \hline
    $\Theta_m=\theta(x),\ \Theta_p=\theta(x+1)$& 0.9999 & 1.0000 & -1.0000 & -0.9999 \\ \hline
    $\Theta_m=\theta(x),\ \Theta_p^{-1}=(e^{-x}+1)$& 0.9998 & 0.9998 &-0.9998  & -0.9997 \\ \hline
   $\Theta_m=\theta(x+1),\ \Theta_p^{-1}=(e^{-x}+1)$& 0.9997 & 0.9998 & -0.9998 & -0.9996 \\ \hline
   \end{tabular}
   \caption{\label{tab:table1}Numerical results for $\mathcal{P}$ for different choices of $\Theta_m(x)$ and $\Theta_p(x)$. We fix $f_\Lambda(n)=e^{-E_n/\Lambda}$, $V(x)=\frac{1}{2} x^2$, and $\Lambda=10^4$. The result shows verifies our statement \eqref{eqn:statement} to high accuracy, consistent with a quantized value for $\mathcal{P}$. We have also tested the deviation from $\pm 1$ decreases as $\Lambda$ increases. }
   \end{table}

   \emph{\color{blue}Topological invariance.--} Having verified \eqref{eqn:statement} for a particular choice of $(f_\Lambda, V, \Theta_p, \Theta_m)$, we now explain its invariance under local continuous deformations. Loosely speaking, the invariance with respect to the deformation of $f_\Lambda$ is a direct consequence of the existence of the limit $\Lambda \rightarrow \infty$, which requires the contribution from states $|n\rangle$ vanishes rapidly enough as $n \rightarrow \infty$. As a result, for two different choices $f_\Lambda$ and $f_\Lambda'$, their difference is peaked near $n\approx \Lambda$, and thus vanishes as we take $\Lambda \rightarrow \infty$. This is consistent with general expectation for regulators, which should not change the underlying physics. More generally, one can take arbitrary initial states in which low-energy Hilbert space is occupied. This includes the thermal equilibrium state with completely different trapping potential $\tilde V(\hat{x})$, where our statement \eqref{eqn:statement} now describe quantized topological response in the quench dynamics \cite{SM}. 

   To understand the invariance of $\mathcal{P}$ for different $(V,\Theta_p, \Theta_m)$, we first imagine the case in which the dimension $d_H$ of the Hilbert space spanned by states $|n\rangle$ is finite. Then, we can safely take the limit of $ \lim_{\Lambda \rightarrow \infty}f_\Lambda(n) =1$, and the L.H.S. of \eqref{eqn:statement} becomes
   $2\pi i~\text{tr}[\hat U^\dagger \Theta_m(\hat{x})\hat U,\Theta_p(\hat{x})]=0.$ Here we have used the cyclic property of the trace operation. As a comparison, our result in \eqref{eqn:statement} is finite for general $t$. The reason is that, without any regulation, both $\text{tr}[\hat U^\dagger \Theta_m(\hat{x})\hat U\Theta_p(\hat{x})]$ and $\text{tr}[\Theta_p(\hat{x})\hat U^\dagger \Theta_m(\hat{x})\hat U]$ are divergent \footnote{As an example, \eqref{eqn:harmonicgen} diverges if we set $\epsilon=0$ at the beginning.}, and it is not possible to use the cyclic property of the trace. (We avoid possible confusion by introducing an explicit regulator in \eqref{eqn:statement}.) This is similar to the derivation of the chiral anomaly \cite{srednicki2007quantum}, and the real-space definition of the 2D Chern number for systems without translation symmetry introduced in Appendix C of \cite{KITAEV20062}. Interestingly, in the latter case, the formula takes a form that is similar to \eqref{eqn:statement}: $(2\pi i)$ times a trace of commutator between asymptotically projective operators. 

   Then, let us consider the difference of $\mathcal{P}$ between two different spatial envelops of the optical impulse $\Theta_p'=\Theta_p+\delta \Theta_p$ and $\Theta_p$.  
    \begin{equation}
    \delta\mathcal{P}=\lim_{\Lambda \rightarrow \infty}2\pi i\sum_n f_\Lambda(n)\langle n| [\hat U^\dagger( \Theta_m(\hat{x})-1/2)\hat U,\delta \Theta_p(\hat{x})]|n\rangle.
    \end{equation}
   Here we have added a $-1/2$ for later convenience, which trivially commutes with arbitrary function. When $\delta \Theta_p(x)$ vanishes rapidly enough at $|x|\rightarrow \infty$, the limit of $\Lambda \rightarrow \infty$ can now be safely taken at first since $\text{tr}[\hat U^\dagger ( \Theta_m(\hat{x})-\frac{1}{2})\hat U\delta \Theta_p(\hat{x})]$ and $\text{tr}[\delta \Theta_p(\hat{x})\hat U^\dagger ( \Theta_m(\hat{x})-\frac{1}{2})\hat U]$ are both finite. This leads to
   \begin{equation}
   \delta \mathcal{P}=2\pi i~\text{tr} [\hat U^\dagger (\Theta_m(\hat{x})-1/2)\hat U,\delta \Theta_p(\hat{x})]=0. 
   \end{equation}
   This shows that a local continuous deformation of the $\Theta_p$ leaves $\mathcal{P}$ invariant. Noticing \eqref{eqn:statement} is symmetric under $\Theta_m \leftrightarrow \Theta_p$ and $\hat U\rightarrow \hat U^\dagger$, we conclude that $\mathcal{P}$ is also invariant under the local continuous deformation of the $\Theta_m$. For the deformation of $V$, we can use
   \begin{equation}
   \delta \hat U(t)=-i\int_0^t dt'~\hat U(t-t') \delta V(\hat x) \hat U(t').
   \end{equation}
   Similar to previous cases, the variation of $\mathcal{P}$ again vanishes when $\delta \hat V(x)$ decays rapidly enough at $|x| \rightarrow \infty$.

   \begin{figure}[t]
   \centering
   \includegraphics[width=0.95\linewidth]{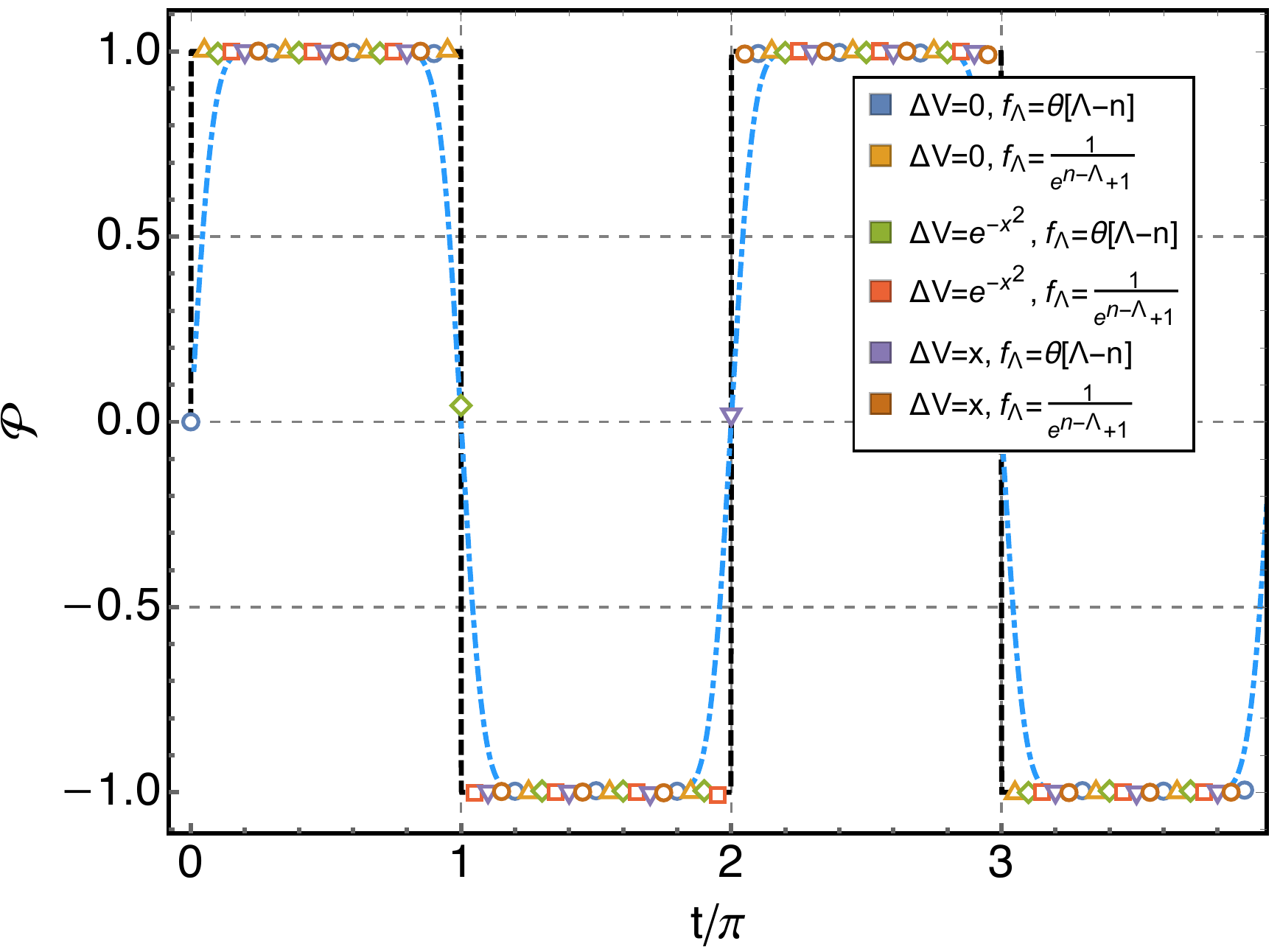}
   \caption{The numerical results of $\mathcal{P}$ for different potential $V(x)=[x^2+\Delta V(x)]/2$ and filling fraction $f_\Lambda(n)$ with finite $\Lambda=150$. The back dashed line is the quantized analytical prediction \eqref{eqn:statement}. The results show deviations of the order of $\sim 10^{-2}$, which is consistent with the quantization of $\mathcal{P}$. The blue dot-dashed is a plot of \eqref{eqn:estimate}, which estimates the realistic effects of finite duration of the optical pulse. Here we set $\sigma =0.2$. }\label{fig2}
   \end{figure}

   Now we present numerical verification of the topological invariance for different choices of $(f_\Lambda,V,\Theta_p,\Theta_m)$. The details of the numerics can be found in the supplementary material \cite{SM}. We first fix $f_\Lambda(n)=e^{-\epsilon E_n}$ and $V(x)=\frac{1}{2} x^2$. The quantized response $\mathcal{P}$ can then be tested to high accuracy by performing the numerically integration in \eqref{eqn:harmonicgen} with small $\epsilon=10^{-4}$. The result is presented in TABLE \ref{tab:table1}, which is consistent with the statement \eqref{eqn:statement} to high accuracy. We then test the invariance of the statement for different choices of the cut-off function $f_\Lambda(n)$ and potential $V(x)$ with fixed $\Theta_m(x)=\Theta_p(x)=\theta(x)$. Since generally, no closed-form expression is available for the Green's function $K(x,y,t)$, we perform an exact diagonalization study in the Hilbert space spanned by the first $L=200$ eigenstates of the harmonic oscillator. Leaving details into the supplementary material \cite{SM}, we present results in FIG \ref{fig2} for $\Lambda=150$. Despite a finite $\Lambda$, the result matches the statement \eqref{eqn:statement} to good accuracy. This guarantees the quantization can be observed in realistic experiments with moderate number of atoms. 
   \vspace{5pt}

   \emph{\color{blue}Practical considerations.--} In realistic experiment the duration of the optical pulse is finite. To estimate the corresponding effect, we make the replacement $$\hat V_p=\xi \Theta_p(\hat{x})\delta(t) \rightarrow \frac{\xi}{\sqrt{2\pi \sigma^2}} \Theta_p(\hat{x})e^{-\frac{t^2}{2\sigma^2}}.$$
   Using the linear response theory, we can determine the density change due to the optical pulse as
   \begin{equation}
\mathcal{P}_\sigma\equiv-2\pi\left.\partial_\xi \rho_\xi(t) \right|_{\xi=0}=\int dt'~\frac{1}{\sqrt{2\pi \sigma^2}}e^{-\frac{t'^2}{2\sigma^2}}\mathcal{P}(t-t').
   \end{equation}
   For a short duration of the pulse $\sigma \omega \ll 1$, $\mathcal{P}_\sigma$ is approximately quantized for $|t- n\pi|\gtrsim \sigma$. The correction of finite $\sigma$ is important near the transition time $t\approx n\pi$. In this case, we can estimate the correction by approximating $\mathcal{P}(t)\approx (-1)^n \text{sgn}(t-n\pi)$, which gives
   \begin{equation}\label{eqn:estimate}
   \mathcal{P}_\sigma\approx(-1)^n \text{erf}\left(\frac{t-\pi  n}{\sqrt{2} \sigma }\right),\ \ \ \ \ \text{for}\ \ |t-\pi  n|\lesssim \sigma.
   \end{equation}
   This describes the smoothen of the response function near $t\approx \pi  n$. A plot of \eqref{eqn:estimate} with $\sigma=0.2$ is presented using the blue dot-dashed line in FIG \ref{fig2}. For $|t- n\pi|\gtrsim \sigma$, it converges to the quantized value with exponentially small corrections. This suggests the quantized response \eqref{eqn:statement} is stable against small time durations of the optical pulse.

   \vspace{5pt}

   \emph{\color{blue}Higher dimensions.--} Finally we study the implication of our statement \eqref{eqn:statement} in $D$-dimensional harmonic traps. The Hamiltonian reads
   \begin{equation}
   H=\sum_{a=1}^DH_a=\sum_{a=1}^D\left(\frac{\hat p_{a}^2}{2}+\frac{\hat x^2_a}{2}\right).
   \end{equation}
   Here we have assumed the trapping frequency $\omega_a=1$. Generalizations to anisotropic harmonic traps is straightforward. Since the $D$-dimensional harmonic trap is exactly analogous to $D$ copies of independent 1D harmonic traps, we propose a straightforward generalization 
   \begin{equation}\label{eqn:statementD}
   \begin{aligned}
   \mathcal{P}^{(D)}&=(2\pi i)^D\lim_{\Lambda \rightarrow \infty}\sum_{\{n_a\}} f_\Lambda\prod_{a=1}^D\langle [\hat U_a^\dagger(t_a) \Theta_m^a(\hat{x}_a)\hat U_a(t_a),\Theta_p^a(\hat{x}_a)]\rangle_{n_a}\\&=\text{sgn}\left(\prod_a\sin t_a\right).
   \end{aligned}
   \end{equation}
   Here $\hat U_a(t)=e^{-i\hat H_a t}$ and $n_a$ is the quantum number in the $x_a$ direction. Both $\Theta_m^a$ and $\Theta_p^a$ satisfies the same boundary condition as their 1D counterparts. We have introduced the regulator $f_\Lambda(\{n_a\})$, which decays rapidly enough for any $n_a\gg \Lambda$. When $f_\Lambda(\{n_a\})=\prod_a f_\Lambda(n_a)$, \eqref{eqn:statementD} is just $D$ copies of the 1D result \eqref{eqn:statement}. We then use the insensitivity of the regulator in the limit of $\Lambda \rightarrow \infty$ to relax the restriction of $f_\Lambda(\{n_a\})$. Numerical verifications is presented in the supplementary material \cite{SM}. When we choose $f_\Lambda(\{n_a\})=(e^{\beta (\sum_an_a-\Lambda)}+1)^{-1}$, the initial state describes a thermal ensemble in $D$-dimension.  
   
   The generalization \eqref{eqn:statementD} can be related to a non-linear response of trapped quantum gases, as an analog of the quantized nonlinear conductance in ballistic metals \cite{PhysRevLett.128.076801}: After preparing the initial state, we add an optical pulse described by the Hamiltonian 
   \begin{equation}
   \hat V_p=\sum_a \xi_a \Theta_p^a(\hat{x})\delta(t-t_a).
   \end{equation}
   We the let the system evolve to time $t_f$, and perform the measurement of $\prod_a\Theta_m^a(\hat x_a)$. For each eigenstate, the contribution from different $a$ factorizes. considering the response in each direction to the linear order, the measurement result $\rho_{\{\xi\}}\equiv\langle \prod_a\Theta_m^a(\hat x_a)\rangle$ satisfies
   \begin{equation}
   (-2\pi)^D\left.\partial_{\xi_1}\partial_{\xi_2}...\partial_{\xi_D} \rho_{\{\xi\}}\right|_{\xi_a=0}=\text{sgn}\left(\prod_a\sin (t_f-t_a)\right).
   \end{equation}
   In particular, for $D=2$, our protocol is reduced to the protocol proposed in \cite{yang2022quantized}, and our result is consistent with the analysis in \cite{yang2022quantized} using semi-classical Boltzmann equation with $\theta_p^a=\theta_m^a=\theta(x)$. Moreover, our analysis in 1D suggest the non-linear response is also topological, regardless of the choices of $(\Theta_p^a,\Theta_m^a)$.
\vspace{5pt}

   \emph{\color{blue}Discussion.--} In this letter, we introduce a universal quantized charge transport of trapped quantum gases. We compute the response function explicitly for a convenient choice of the setup, and show the result is topological invariant under local continuous deformations with fixed boundary conditions for the trapping potential $V$, atom distribution $f_\Lambda$, the spatial envelope of the optical pulse $\Theta_p$, and the measurement region $\Theta_m$. The statement is supported by various numerical results, which matches the analytical prediction to high accuracy. After analyzing realistic effects in experiments, we believe our statement \eqref{eqn:statement}, as well as its higher-dimensional generalization \eqref{eqn:statementD} for non-linear responses, can be directly observed in the near-term experiments using ultracold atomic gases. 

\vspace{5pt}

\textit{Acknowledgment.} We thank Yingfei Gu, Chengshu Li, Ning Sun, and Fan Yang for invaluable discussions. PZ acknowledges support from the Walter Burke Institute for Theoretical Physics at Caltech.

\bibliography{draft.bbl}

\onecolumngrid
\begin{center}
\newpage\textbf{\large
Supplementary Material: Quantized Topological Response in Trapped Quantum Gases}
\\
\vspace{4mm}

\end{center}

In this supplementary material, we present results for: 1. The derivation of  Eq. (6); 2. The numerical details; 3. The numerical verification of the quantized response in quantum quenches; 4. The numerical verification in 2D.

\section{The derivation of Eq. (6)}

In this section, we present details for the derivation of the Eq. (6) in the main text. After taking $f_\Lambda(n)=e^{-E_n/\Lambda}$, $V(x)=\frac{1}{2} x^2$, and $\Theta_p(x)=\Theta_m(x)=\theta(x)$, the Eq. (5) becomes 
\begin{equation}
\begin{aligned}
&\mathcal{P}=4\pi\lim_{\epsilon \rightarrow 0}\text{Im}\int_0^\infty dx\int_0^\infty dy~K(x,y,t-i\epsilon)K(y,x,-t),
\\&K(x,y,t)=\frac{1}{\sqrt{2\pi i \sin t}}\exp(\frac{i}{2\sin t}\Big[(x^2+y^2)\cos t-2xy\Big]).& 
\end{aligned}
\end{equation}
Since the factor $1/{\sqrt{2\pi i \sin t}}$ contains no $x$ or $y$ variable, we only need to compute
\begin{equation}
\mathcal{I}=\int_0^\infty dx\int_0^\infty dy~\exp(\frac{i}{2\sin t_1}\Big[(x^2+y^2)\cos t_1-2xy\Big])\exp(-\frac{i}{2\sin t_2}\Big[(x^2+y^2)\cos t_2-2xy\Big]).
\end{equation}
The integration over $x$ can be computed by using the error function:
\begin{equation}
\mathcal{I}=-\int_0^\infty dy\frac{e^{\frac{5}{4}\pi i} \sqrt{\frac{\pi }{2}} e^{-i y^2 \tan
   \left(\frac{t_1-t_2}{2}\right)} }{\sqrt{\cot (t_1)-\cot (t_2)}}\left(1-\text{erf}\left(\frac{e^{\frac{3}{4}\pi i}
   y (\csc (t_1)-\csc (t_2))}{\sqrt{2} \sqrt{\cot (t_1)-\cot
   (t_2)}}\right)\right).
\end{equation}
There are two terms in the bracket. The first term is symmetry under the reflection $y\rightarrow -y$, and does not contributes to the non-trivial charge transport. Keeping the seond term only, we find
\begin{equation}
\begin{aligned}
&\mathcal{\tilde{I}}=-\frac{\left(\frac{1}{2}+\frac{i}{2}\right) (-1)^{3/4} (\csc (t_1)-\csc
   (t_2)) }{\sqrt{i \tan
   \left(\frac{t_1-t_2}{2}\right)} (\cot (t_1)-\cot (t_2))
   \eta(t_1,t_2)}\tan ^{-1}\left(\frac{\eta(t_1,t_2)}{\sqrt{2} \sqrt{i \tan
   \left(\frac{t_1-t_2}{2}\right)}}\right),\ \ \ \ \ \ \eta(t_1,t_2)=\sqrt{-\frac{i (\csc (t_1)-\csc
   (t_2))^2}{\cot (t_1)-\cot (t_2)}}.
   \end{aligned}
\end{equation}
Now we can substitue $t_1=t-i\epsilon$ and $t_2=t$. After a Taylor expansion, we find the Eq. (6) in the main text
\begin{equation}
\mathcal{P}=\sqrt{\sin ^2(t)} \csc
   (t)+\frac{\epsilon ^2 (\cos (2 t)-7) \csc (t)}{48 \sqrt{\sin ^2(t)}}+O(\epsilon^4)=\text{sgn}\left(\sin(t)\right)\left[1+\frac{\epsilon ^2 (\cos (2 t)-7) }{48 \sin^2(t)}+O(\epsilon^4)\right].
\end{equation}
This is valid for $t\neq n\pi$, as mentioned in the main text.

  \section{The details of numerical calculations}
Now we present numerical details for the TABLE 1 and FIG 2 presented in the main text. For the TABLE 1, we fix $f_\Lambda(n)=e^{-E_n/\Lambda}$, $V(x)=\frac{1}{2} x^2$ while tuning $\Theta_p$ and $\Theta_m$. The result of $\mathcal{P}$ can then be obtained by performing the numerical integral 
\begin{equation}
\begin{aligned}
   &\mathcal{P}=4\pi\lim_{\epsilon \rightarrow 0}\text{Im}\int_{-\infty}^{\infty} dx\int_{-\infty}^{\infty} dy~\Theta_m(x)\Theta_p(y)K(x,y,t-i\epsilon)K(y,x,-t),
   \\&K(x,y,t)=\frac{1}{\sqrt{2\pi i \sin t}}\exp(\frac{i}{2\sin t}\Big[(x^2+y^2)\cos t-2xy\Big]).& 
   \end{aligned}
\end{equation}

For FIG 2, we fix $\Theta_p(x)=\Theta_m(x)=\theta(x)$ and study the problem using exact diagonalization. For a general single-particle Hamiltonian
\begin{equation}
\hat H=\frac{\hat p^2+\hat x^2}{2}+\Delta V(\hat x).
\end{equation}
We present the Hamiltonian using the annihilation operator $\hat x=\frac{1}{\sqrt{2}}(\hat a^\dagger+\hat a)$ $\hat x=\frac{i}{\sqrt{2}}(\hat a^\dagger -\hat a)$:
\begin{equation}
\hat H=\left(\hat a^\dagger \hat a+\frac{1}{2}\right)+\Delta V\left(\frac{1}{\sqrt{2}}(\hat a^\dagger+\hat a)\right).
\end{equation}
In numerics, we use the eigenstate of harmonic oscillator $\hat a^\dagger \hat a|n\rangle_h=n|n\rangle_h$, with a cutoff of $n=1,2,...,L$. We set $L=200$ for FIG 2. The numerical diagonalization gives the eigenstate of $\hat H$ with eigenenergy $E_n$ as $|n\rangle=\sum_m U_{nm}|m\rangle_h$. We have
\begin{equation}\label{eqn:numsum}
\begin{aligned}
    \mathcal{P}=&\lim_{\Lambda \rightarrow \infty}2\pi i\sum_n f_\Lambda(n)\langle n| [\hat U^\dagger \theta(\hat{x})\hat U,\theta(\hat{x})]|n\rangle
    =-\lim_{\Lambda \rightarrow \infty}4\pi  ~\text{Im}\sum_n f_\Lambda(n)\langle n| \hat U^\dagger \theta(\hat{x})\hat U\theta(\hat{x})|n\rangle\\
    =&-\lim_{\Lambda \rightarrow \infty}4\pi ~\text{Im}\sum_{nm}f_\Lambda(n) e^{i(E_n-E_m)t} \langle n|\theta(\hat{x})|m\rangle \langle m|\theta(\hat{x})|n\rangle.
    \end{aligned}
    \end{equation} 
Here the matrix element takes the form
\begin{equation}\label{eqn:numsum2}
\langle m|\theta(\hat{x})|n\rangle=\sum_{m'n'}U^\dagger_{mm'} \ _h\langle m'|\theta(\hat{x})|n'\rangle_hU_{n'n},
\end{equation}
$\ _h\langle m|\theta(\hat{x})|n\rangle_h$ can be computed using Hermite polynomials. Only $m\neq n$ terms contribute to \eqref{eqn:numsum}, which is non-zero only if $m+n$ is odd. The result reads
\begin{equation}\label{eqn:theta}
\ _h\langle m|\theta(\hat{x})|n\rangle_h=\sum_{l=0}^{\text{min}(m,n)}\frac{  2^{-l+m/2+n/2-1}l!}{\sqrt{m! n! } \Gamma
   \left(l-\frac{m}{2}-\frac{n}{2}+1\right)}\binom{m}{l} \binom{n}{l}\ \ \ \ \ \ \ \ \ \ \ \ (\text{odd}\ m+n).
\end{equation}
Using \eqref{eqn:numsum2} and \eqref{eqn:theta}, $\mathcal{P}$ can be computed efficiently, which gives the result shown in FIG 2 for different $\Delta V$ and $f_\Lambda$.

 \begin{figure}[t]
   \centering
   \includegraphics[width=0.45\linewidth]{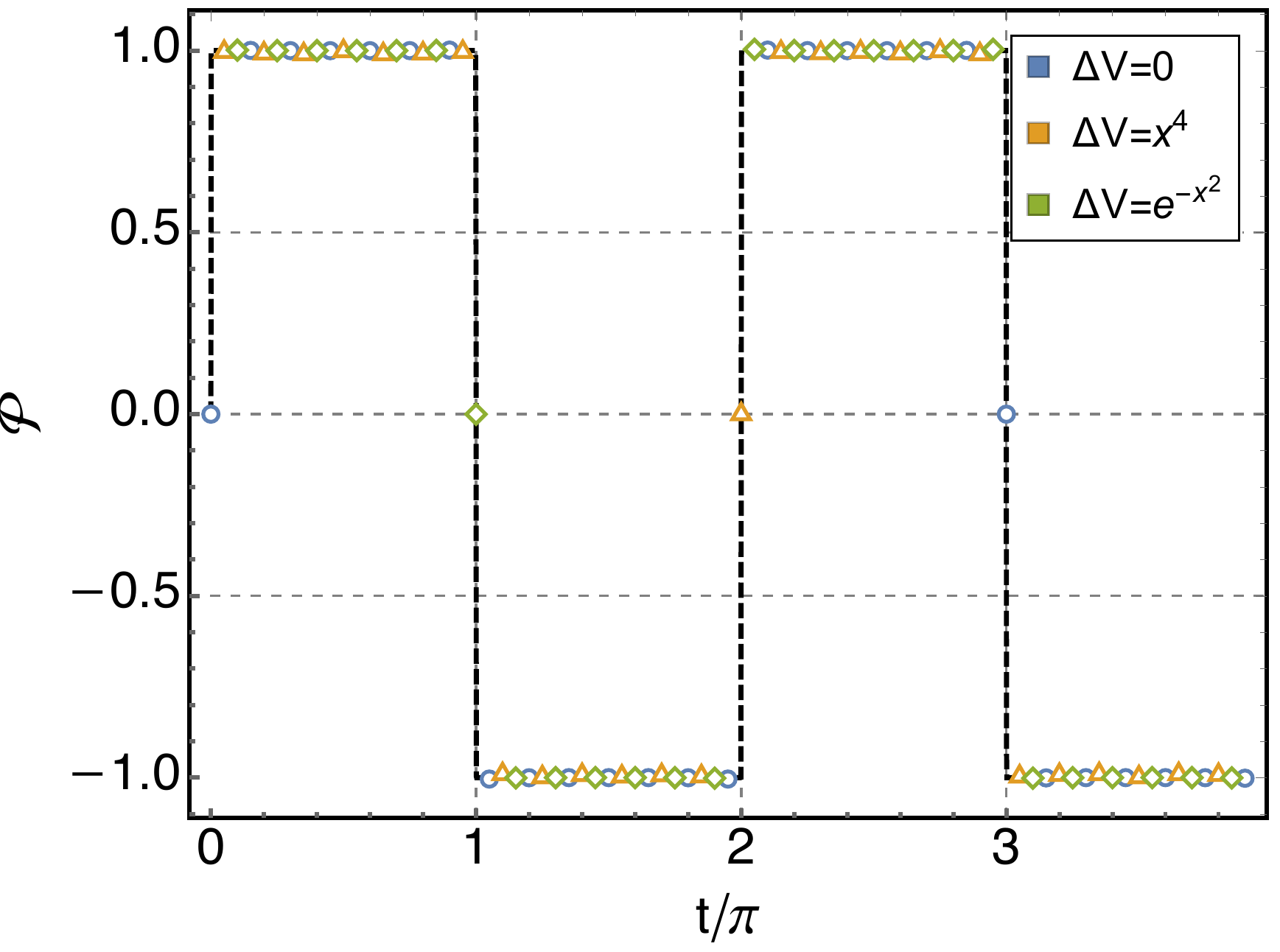}
   \caption{The numerical results of $\mathcal{P}$ for different initial potential $\tilde V(x)=[x^2+\Delta V(x)]/2$ in the quench dynamics with evolution trapping potential $V=x^2$ and $f_\Lambda=(e^{\tilde{E}_n-80}+1)^{-1}$. The result is consistent with $\mathcal{P}=\text{sgn}(\sin t)$ to good precision. }\label{supp1fig}
   \end{figure}
\section{More numerical verifications of the universal quantized topological response}

\subsection{Quantized response in quench dynamics}
Now we present numerical verification of the quantized response in quantum quenches. Experimentally, this corresponds to prepare the initial state with respect to the Hamiltonian $\hat{\tilde{H}}$:
\begin{equation}
\hat{\tilde{H}}=\frac{\hat p^2}{2m}+\tilde{V}(\hat{x}).
\end{equation}
Here we still assume all eigenstates are bounded, but without any restriction for the asymptotic behavior of $\tilde{V}(x)$. As an example, one may take a quartic potential $\tilde{V}(x)=x^4/2$. Similar to the original protocol in the main text, we prepare an initial state where $|\tilde{n}\rangle$ (eigenstate of $\hat{\tilde{H}}$ with energy $\tilde{E}_n$) is filled with probability $f_\Lambda(\tilde{n})$, and then apply the optical pulse. The system is then evolved under a different Hamiltonian $\hat{H}$ for time $t$, in which the trapping potential is asymptotically quadratic. Since the initial state is not an eigenstate of the evolution Hamiltonian, this describes a quantum quench. Finally, the density of the system is measured as in the main text. As discussed in the main text, the necessary condition for the quantized response is that the initial state should project out the Hilbert space at very high energy. Consequently, the details of the initial trapping potential are not important.

The numerical verification of above arguments is shown in FIG \ref{supp1fig}. Here we consider different initial states with $\tilde{V}=(x^2+\Delta V)/2$. Here we fix $f_\Lambda=(e^{n-\Lambda}+1)^{-1}$, with $\Lambda=80$, and evolution trapping potential $V=x^2/2$. We use a relative small $\Lambda$ since for $\Delta V\sim x^4$ the eigenstates will receive large finite $L$ effect in the numerical diagonalization method. Nevertheless, the result matches with the analytical prediction $\mathcal{P}=\text{sgn}(\sin t)$ to good precision.

 \begin{figure}[t]
   \centering
   \includegraphics[width=0.85\linewidth]{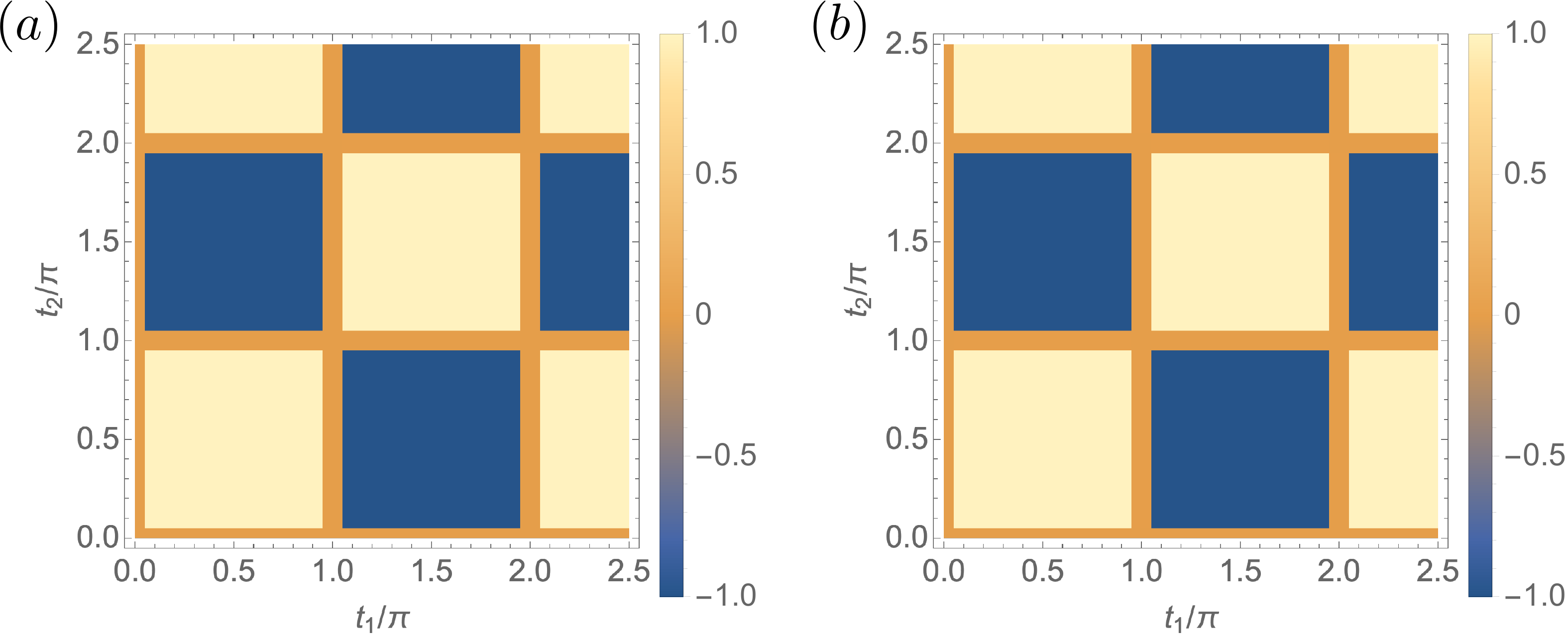}
   \caption{Density plots of $\mathcal{P}$ with steps $\delta t=0.1\pi$ in 2D harmonic traps ford (a) $f_\Lambda(n_1,n_2)=(1+e^{n_1+n_2-2\Lambda})^{-1}$ and (b). $f_\Lambda(n_1,n_2)=(1+e^{n_1-\Lambda})^{-1}(1+e^{n_2-\Lambda})^{-1}$.  with $\Lambda=80$. The result shows the quantization of $\mathcal{P}^{(2)}$ is independent of the cutoff function $f_\Lambda(n_1,n_2)$. }\label{supp2fig}
   \end{figure}

\section{The numerical verification for the regulator independence in 2D }
In this subsection, we provide numerical verifications of replacing $\prod_a f_\Lambda(n_a)$ with general regulator $f_\Lambda(\{n_a\})$ in higher-dimensional generalizations. In the calculation, we fix $V=x^2/2$ with $\tilde{V}=V$. Similar to the 1D expansion \eqref{eqn:numsum}, the response function in 2D takes the form
\begin{equation}\label{eqn:numsum}
\begin{aligned}
    \mathcal{P}^{(2)}=\lim_{\Lambda \rightarrow \infty}16\pi^2 ~\sum_{n_1n_2}\sum_{m_1m_2}f_\Lambda(n_1,n_2)\prod_{a=1}^2 \left[\sin({(n_a-m_a)t_a}) \langle n_a|\theta(\hat{x}_a)|m_a\rangle \langle m_a|\theta(\hat{x}_a)|n_a\rangle\right].
    \end{aligned}
    \end{equation} 
    The summation can be computed numerically using \eqref{eqn:numsum2}. The result for different $f_\Lambda(n_1,n_2)$ is shown in FIG \ref{supp2fig} with $\Lambda=150$, which verifies the invariance of $\mathcal{P}^{(2)}$ under continuous deformations of $f_\Lambda(n_1,n_2)$.
	
\end{document}